\documentclass[reprint,showpacs,aps,prl,twocolumn,amsmath,amssymb]{revtex4}

\usepackage[dvipdfmx]{graphicx}
\usepackage[dvipdfmx]{color}
\usepackage{url}
\usepackage{dcolumn}
\usepackage{bm}
\usepackage[dvipdfmx,bookmarkstype=toc=true,colorlinks=true,urlcolor=blue,%
      linkcolor=blue,citecolor=blue,linktocpage=true,bookmarks=false]{hyperref}
\usepackage{xspace}
\newcommand{\FeTeSe}{FeTe$_{0.6}$Se$_{0.4}$\xspace}

\begin{document}

\title{Evidence for a cos(4$\varphi$) Modulation of the Superconducting Energy Gap of Optimally Doped FeTe$_{0.6}$Se$_{0.4}$ Single Crystals Using Laser Angle-Resolved Photoemission Spectroscopy}

\author{K.~Okazaki$^{1}$}
\email{okazaki@issp.u-tokyo.ac.jp}
\author{Y.~Ito$^{1}$}
\author{Y.~Ota$^{1}$} 
\author{Y.~Kotani$^{1}$}
\author{T.~Shimojima$^{2}$} 
\author{T.~Kiss$^{3}$}
\author{S.~Watanabe$^{4}$}
\author{C.~-T.~Chen$^{5}$} 
\author{S.~Niitaka$^{6,7}$} 
\author{T.~Hanaguri$^{6,7}$} 
\author{H.~Takagi$^{6,7}$}
\author{A.~Chainani$^{8}$}
\author{S.~Shin$^{1,7,8,9}$}
\affiliation{
$^{1}$Institute for Solid State Physics (ISSP), University of Tokyo, Kashiwa, Chiba 277-8581, Japan\\
$^{2}$Department of Applied Physics, University of Tokyo, Tokyo 113-8656, Japan\\
$^{3}$Graduate School of Engineering Science, Osaka University, Osaka 560-8531, Japan\\
$^{4}$Research Institute for Science and Technology, Tokyo University of Science, Chiba 278-8510, Japan\\
$^{5}$Beijing Center for Crystal R\&D, Chinese Academy of Science (CAS), Zhongguancun, Beijing 100190, China\\
$^{6}$RIKEN Advanced Science Institute, 2-1, Hirosawa, Wako, Saitama 351-0198, Japan \\
$^{7}$TRIP, JST, Chiyoda-ku, Tokyo 102-0075, Japan\\
$^{8}$RIKEN SPring-8 Center, Sayo-gun, Hyogo 679-5148, Japan\\
$^{9}$CREST, JST, Chiyoda-ku, Tokyo 102-0075, Japan
}

\date{\today}

\begin{abstract}
We study the superconducting-gap anisotropy of the $\Gamma$-centered hole Fermi surface in optimally doped \FeTeSe ($T_c$ = 14.5 K), using laser-excited angle-resolved photoemission spectroscopy (ARPES). We observe sharp superconducting (SC) coherence peaks at $T$ = 2.5 K. In contrast to earlier ARPES studies but consistent with thermodynamic results, the momentum dependence shows a $\cos(4\varphi)$ modulation of the SC-gap anisotropy. 
The observed SC-gap anisotropy strongly indicates that the pairing interaction is not a conventional phonon-mediated isotropic one. Instead, the results suggest the importance of second-nearest-neighbor electronic interactions between the iron sites in the framework of $s_\pm$-wave superconductivity.
\end{abstract}

\maketitle

\noindent
In contrast to the single band or orbital ($3d_{x^2-y^2}$) character of electronic states in the high-temperature cuprates, the complex properties of the Fe-based superconductors (FeSCs)~\cite{Kamihara2008JACS} originate in its multiorbital or multiband electronic structure. The multiband nature of the FeSCs manifests typically in the form of hole Fermi surfaces (FSs) around the Brillouin-zone center ($\Gamma$ point) and electron FSs around the zone corner ($M$ point in the primitive tetragonal unit cell)~\cite{Singh2008PRL,Singh2008PRB}. Theoretical studies
have shown that the multiple FS sheets can give $s$-wave superconductivity with a sign reversal between the hole and electron FS sheets (an $s_\pm$-wave state), arising from Cooper pair formation mediated by spin fluctuations~\cite{Mazin2008PRL,Kuroki2008PRL}. On the other hand, the multiorbital nature can also induce orbital fluctuations possibly enhanced by electron-phonon interactions, and it was proposed that orbital fluctuations can give rise to $s$-wave superconductivity with no-sign-reversal(an $s_{++}$-wave state)~\cite{Kontani2010PRL}. 

It is difficult to distinguish between $s_\pm$-wave and $s_{++}$-wave states because it requires a determination of the phase of the order parameter. The observation of half-integer quantum flux jumps in a Nb-NdFeAsO$_{0.88}$F$_{0.12}$ loop has been interpreted as a $\pi$ phase shift of the Josephson current and this should occur only when the sign reversal exists as in the $s_\pm$-wave state~\cite{Chen2010NP}. Other evidence for $s_\pm$ wave has been found in the Fe(Te,Se) (abbreviated as 11) system : it was shown that the magnetic-field dependence of the quasiparticle interference pattern observed by scanning tunneling spectroscopy is consistent with a sign-reversal $s_\pm$-wave state~\cite{Hanaguri2010Science}. However, the robustness of $T_c$ against impurity doping is difficult to explain in terms of an $s_\pm$-wave state~\cite{Sato2010JPSJ,Lee2010JPSJ}, and thus, it supports a no-sign-reversal $s_{++}$-wave state indicating importance of orbital fluctuations~\cite{Kontani2010PRL}. Observation of an orbital nematic state in the underdoped compounds of FeSCs~\cite{Chuang2010Science} may also indicate the importance of orbital fluctuations in the FeSCs. 
Hence, further verifications is needed regarding the importance of spin and orbital fluctuations in each system of FeSCs.

The SC-gap structure of the 11 system has been studied by magnetic penetration depth~\cite{ Kim2010PRB,Serafin2010PRB,Bendele2010PRB}, thermal conductivity~\cite{Dong2009PRB}, nuclear magnetic resonance~\cite{Michioka2010PRB}, and field angle resolved specific heat (ARSH)~\cite{Zeng2010NC} measurements. All these measurements indicate a spin singlet superconductivity with anisotropic gaps but no nodes for this system. 
In this Letter,
we report detailed SC-gap measurements of the optimally doped ``11" system, \FeTeSe ($T_c$ $\sim$ 14.5 K), using laser-excited angle-resolved photoemission spectroscopy [laser angle-resolved photoemission spectroscopy (ARPES)]. Among the various FeSCs, the 11 system has the simplest crystal structure consisting of stacked layers of Fe(Te,Se)$_4$ tetrahedra only. From this simplicity of the crystal structure, the cleaved surfaces are not polar surfaces as in the ``1111" and ``122" systems but are charge-neutral surfaces. Based on this aspect and combined with the bulk sensitivity of laser ARPES~\cite{note_EscapeDepth}, it is expected
that we unveil the intrinsic symmetry of the superconducting order parameter. Our results do reveal a SC-gap structure with a clear $\cos(4\varphi)$ modulation.  
While some SC-gap anisotropies have been reported by the ARPES measurements of ``122" and ``111" systems~\cite{Xu2011NP,Borisenko2012Symmetry}, the previous ARPES studies of the 11 system indicated an isotropic SC gap~\cite{Nakayama2010PRL,Miao2012PRB}. Our results seem to be in strong contradiction to these results. However, our results are quite consistent with the ARSH study in a rotating magnetic field that concluded a significant gap anisotropy with gap minima along the $\Gamma$-$M$ direction, i.e. along the nearest neighbor Fe-Fe bond direction~\cite{Zeng2010NC}. 
We discuss the origin of the $\cos(4\varphi)$ modulation from the aspects of the spin and orbital fluctuations, and the importance of second-nearest-neighbor electronic interactions between the iron sites is suggested in the framework of $s_\pm$-wave superconductivity.

Single crystals of FeTe$_{0.6}$Se$_{0.4}$ were prepared by a melt-growth technique. Chemical composition of the grown crystals was determined by electron probe microanalysis and inductively coupled plasma atomic emission spectrometry.  The samples 
showed sharp superconducting transitions at $T_c$ = 14.5 K, confirming the optimal doping. Details have been described in Ref.~\onlinecite{Hanaguri2010Science}.
ARPES data were collected using a laser-ARPES apparatus developed at ISSP with the VG-Scienta HR8000 electron analyzer and the 6.994 eV quasi-CW laser (repetition rate = 120 MHz)~\cite{Okazaki2012Science}. The overall energy resolution was set to $\sim$ 1.2 meV and the angular resolution was 0.1 deg. The Fermi edge of an evaporated gold film was measured to calibrate $E_F$ energy positions. 
Band-structure calculations were carried out using the Wien2k code for the parent FeTe. The lattice parameters were taken from those obtained by powder neutron diffraction measurements~\cite{Li2009PRB}, as in a previous report~\cite{Miyake2010JPSJ}. We confirmed that the obtained band dispersions were in accord with the earlier study~\cite{Miyake2010JPSJ}.


\begin{figure} [t]
\begin{center}
\includegraphics[scale=0.52]{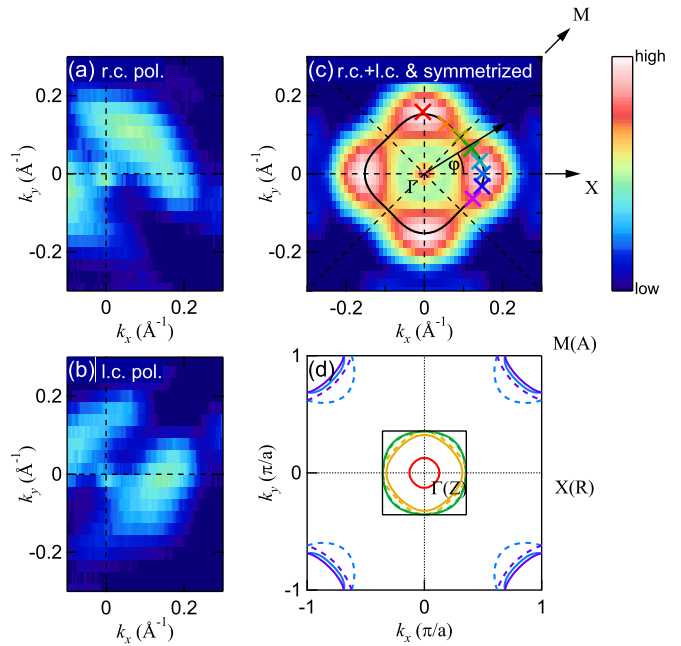} 
\caption{(a), (b) FS intensity map of FeTe$_{0.6}$Se$_{0.4}$ measured at 25 K with right- and left-circularly polarized light, respectively. The integration energy window is $\pm$ 5 meV from $E_F$. (c) FS map image obtained by summing up the spectra in (a) and (b) and symmetrized by taking into account the tetragonal crystal symmetry. The symbols indicate the $k_F$ positions where the SC gap was measured. The solid line corresponds to the FS shape deduced from fitting the $k_F$ positions to the model function with fourfold symmetry. The definition of FS angle $\varphi$ is indicated. (d) FSs of the pure FeTe given by the band-structure calculation. Solid and dashed lines indicate those at $k_z$ = 0 ($\Gamma$-$X$-$M$) plane and $k$ = $\pi$ ($Z$-$R$-$A$) plane, respectively. The innermost FS around the $\Gamma$ point is closed around this point and does not exist around the $Z$ point due to its three dimensionality. The other two FSs around the $\Gamma$ point are nearly two dimensional. The solid square corresponds to the region displayed in (c)~\cite{note_k}. }
\label{Fig1} 
\end{center}
\end{figure}

Figures 1(a) and (b) show the FS maps of \FeTeSe measured at 25 K with right- and left-circularly polarized light, respectively. Figure 1(c) shows the FS map image obtained by summing up the spectra measured with right- and left-circularly polarized light in order to decrease matrix element effects and symmetrized by taking into account the tetragonal crystal symmetry. We confirmed that the band dispersions determined from the peak positions of the second derivative of energy distribution curve (EDC) spectra are basically consistent with reported band dispersions~\cite{Chen2010PRB,Tamai2010PRL,Nakayama2010PRL}. In particular, we confirmed that the laser ARPES results probe the $k_{z} \sim 0$ plane and shows that instead of the three hole FSs expected from the band-structure calculations~\cite{Subedi2008PRB,Miyake2010JPSJ}, the experimental data shows that only one hole band crosses the $E_F$. We determined the $k_F$ positions for several cuts from the peak positions of the momentum distribution curves. The results are indicated by the symbols in Fig. 1(c). The solid line in Fig. 1(c) indicates the FS shape deduced from fitting the $k_F$ positions to a model function with fourfold symmetry. 
The fit shows that the FS is anisotropic and the FS area is reduced in comparison with that of the calculated one for the parent FeTe [Fig. 1(d)]. This may indicate that the Se substitution effectively leads to a reduction in the hole FS area or may be originated from correlation effects~\cite{Yin2011NM}.

Figure 2(a) shows the EDCs at $k_F$ measured below $T_c$ (2.5 K) and above $T_c$ (25 K). To cancel out the effect of the Fermi-Dirac cutoff, the EDCs were symmetrized with respect to $E_F$, and the results are shown in Fig. 2(b). Each EDC is identified with a FS angle $\varphi$ indicated in Fig. 2(b). Sharp superconducting coherence peaks can be recognized very clearly in the spectra below $T_c$. The vertical dashed line indicates the peak position of the EDC at $\varphi$ = 44.5$^\circ$. The EDCs at $\varphi$ = 0.3$^\circ$ and 89.5$^\circ$, for example, have larger peak energies indicating a finite SC-gap anisotropy. In order to quantify the SC-gap sizes, we fitted the spectra to the BCS spectral function~\cite{Matsui2003PRL,Shimojima2011Science} and the results are shown as the solid lines in Fig. 2. 
The observed superconducting coherence peaks are well reproduced by the fitting function, and thus the SC-gap sizes can be evaluated quantitatively. 

\begin{figure} [t]
\begin{center}
\includegraphics[scale=0.48]{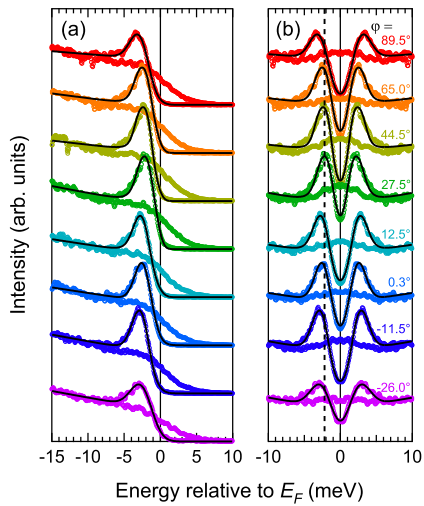} 
\caption{ (a) EDCs and (b) symmetrized EDCs with respect to $E_F$ at $k_F$ measured below $T_c$ (2.5 K) and above $T_c$ (25 K). To cancel out the effect of the Fermi-Dirac cutoff, these EDCs are symmetrized with respect to $E_F$, and the results are shown in Fig. 2(b). Color of each EDC and symmetrized EDC corresponds to that of the $k_F$ positions in Fig. 1(c). The FS angle of each EDC is indicated in Fig. 2(b). The solid lines are the fitting functions using the BCS spectral function.}
\label{Fig2}
\end{center}
\end{figure}

\begin{figure} [b]
\begin{center}
\includegraphics[scale=0.4]{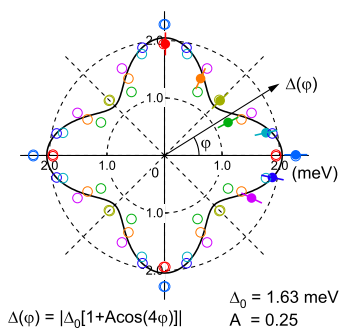} 
\caption{FS-angle dependence of the SC-gap size obtained from fitting to the BCS spectral function. Solid circles are the SC-gap sizes obtained from fitting of the spectra in (a) and open circles are symmetrized by taking into account the tetragonal symmetry of the structure.  Color of each symbol corresponds to that of the $k_F$ positions in Fig. 1(c). The solid line is a fitting result using the model SC-gap function $\Delta(\varphi) = |\Delta\left[1+A\mathrm{cos}(4\varphi)\right]|$, where $\varphi$ is the FS angle. 
}
\label{Fig3}
\end{center}
\end{figure}

In Fig. 3, the SC-gap sizes obtained from the fitting are plotted in a polar graph as solid circles. The error bars of the estimated SC-gap sizes are $\pm$ 0.2 meV as indicated by solid bars. The open circles are symmetrized by taking into account the tetragonal crystal symmetry. In spite of the finite error bars of the estimated SC-gap sizes, a clear anisotropy of the SC-gap size can be recognized and it obeys the tetragonal crystal symmetry. We have fitted the SC-gap sizes to a model anisotropic $s$-wave function with fourfold symmetry,
\[
\Delta(\varphi) = |\Delta\left[1+A\mathrm{cos}(4\varphi)\right]|.
\]
The solid line indicates the fitting result and well reproduces the observed SC-gap anisotropy. The obtained fitting values are the average SC-gap size $\Delta_0$ = 1.63 meV and 25 \% $\cos(4\varphi)$ modulation. This corresponds to the maximum ($\Delta_{max}$) and minimum ($\Delta_{min}$) SC-gap sizes of 2.04 and 1.22 meV, respectively, and the $2\Delta_{max}/k_BT_c$ and  $2\Delta_{min}/k_BT_c$ of 3.27 and 1.95, respectively. We conclude that a large anisotropy exists in the SC gap of FeTe$_{0.6}$Se$_{0.4}$ in the form of a cos(4$\varphi$) modulation, in strong contradiction to previous studies~\cite{Nakayama2010PRL,Miao2012PRB}. This most probably originates in the bulk sensitivity and energy resolution of the present laser-ARPES measurements compared to the earlier more surface sensitive measurements. However, it is noted that the difference between the present and previous studies could also arise from the existence of excess iron i.e. nonstoichiometry, which is known to play a notorious role in the ``11" system~\cite{Li2009PRB}. For our samples, we accordingly confirmed that excess iron could not be detected within the error bars of electron probe microanalysis and inductively coupled plasma atomic emission spectroscopy and that the number of bright spots in the STM topographic image, which is assigned to excess iron atoms, was only 0.1 \% of the total number of iron atoms~\cite{Hanaguri2010Science}. 
More importantly, since the observed gap anisotropy is consistent with the results of ARSH measurements, we believe our study does indeed reveal the intrinsic symmetry of the SC gap. However, it is noted that our results indicate a lower bound of the SC-gap modulation and it could be as large as 50 \% based on the ARSH results. This may be due to effects of disorder that tend to homogenize the gap, but which are inevitable for substituted systems. Further, it is known that even without excess Fe atoms, the 11 systems show higher residual resistivity ($>$ 100 $\mu\Omega$cm) compared to the Ba122(As/P), LaFePO, LiFeAs and KFe$_2$As$_2$ systems~\cite{Serafin2010PRB}. While our samples have a very small amount of excess Fe atoms, they show optimal $T_c$ and in combination with the higher bulk sensitivity of laser ARPES~\cite{note_EscapeDepth}, it allowed us to investigate the anisotropy of the gap.
It is also noted that the ARSH study attributed the anisotropy primarily to the zone corner ($M$ point)-centered electron FS, and only secondarily to the $\Gamma$-centered hole FS~\cite{Zeng2010NC}. 
We would like to note also that the average SC-gap size $\Delta_0$ = 1.63 meV is quite consistent that a clear peak was observed at $\sim \pm$ 1.7 meV in the tunneling spectra~\cite{Hanaguri2010Science}. Also, it is interesting that the SC-gap size is larger at the FS angle where the spectral intensity at $E_F$ is more intense (Fig. 1(c)). This probably means that states with larger density of states at $E_F$ have larger SC-gap sizes.

Based on the observed $\mathrm{cos}(4\varphi)$ modulation for the SC-gap anisotropy and comparison with thermodynamic studies, it is clear that the pairing interactions are neither of the $d_{x^2-y^2}$ type 
nor of the conventional phonon-mediated isotropic $s$-wave type. Recently, Maiti {\it et al.} succeeded in explaining the origin of the SC-gap nodes of KFe$_2$As$_2$~\cite{Maiti2012PRB}
observed by laser ARPES~\cite{Okazaki2012Science}. They used a simple model with an angle-dependent interband and intraband electronic interactions. Their model can also reproduce the spin-fluctuation mediated $s_{\pm}$ superconductivity when both hole and electron FS sheets are present~\cite{Chubukov2009PRB}. According to their model, the SC-gap anisotropy can originate from the elongation of the FS shape and anisotropic electronic interactions. In contrast, if the electronic interactions are fully isotropic, the SC-gap anisotropy should be anticorrelated with the elongation of the corresponding FS~\cite{Maiti2012PRB}. However, the directions of the observed SC-gap maxima coincide with the FS elongation, i.e., along the $\Gamma$-$X$ directions, the FS is elongated, the SC gaps show maxima. This requires anisotropic interactions and indicates that the $(\cos k_x + \cos k_y)$-type interaction is important in the 11 system. Because the nearest-neighbor irons are located at ($\pm a/2$,$\pm a/2$) and the second-nearest-neighbor irons are located at ($\pm a$,0) and (0,$\pm a$), the interactions between the nearest-neighbor and second-nearest-neighbor irons can be described as 
\begin{eqnarray*}
e^{i(k_x+k_y)a/2}+e^{i(k_x-k_y)a/2}+e^{i(-k_x+k_y)a/2}+e^{i(-k_x-k_y)a/2} \\
= 4\cos(k_xa/2)\cos(k_ya/2)
\end{eqnarray*}
and 
\begin{eqnarray*}
e^{ik_xa}+e^{-ik_xa}+e^{ik_ya}+e^{-ik_ya} = 2[\cos(k_xa)+\cos(k_ya)],
\end{eqnarray*}
respectively. Thus, the $(\cos k_x + \cos k_y)$-type interaction originates from next-nearest-neighbor iron sites, and may be related to the fact that the parent compound of the 11 system has a double-stripe antiferromagnetic structure different from other FeSCs with single-stripe antiferromagnetic structures~\cite{Li2009PRB}.

On the other hand, the SC-gap anisotropy can be expected also from the orbital fluctuation theory~\cite{Kontani2010PRL}. It is determined by the ``interorbital nesting" between the $x^2-y^2$ and $xz$ of $yz$ orbitals~\cite{Kontani_PC}. Because the contribution of the $x^2-y^2$ orbital is dominant to the hole FS sheet whose SC gaps were measured, depending on the contribution of $xz$ or $yz$ orbitals to the electron FS sheets, the observed SC-gap anisotropy can be reproduced by the ``interorbital nesting".  Anyway, our observation from the bulk sensitive and high-resolution measurements revealed a detailed and intrinsic nature of the SC-gap anisotropy and would impose strong restrictions on the theories.

In summary, we have observed SC gap anisotropy in the $\Gamma$ centered hole FS of \FeTeSe by laser ARPES. A $\cos(4\varphi)$ modulation was observed in the SC-gap anisotropy very clearly and the SC-gap maxima coincides with the FS elongation along $\Gamma$-$X$.
The observed momentum dependence is in contrast to earlier ARPES studies but consistent with thermodynamic results. 
The observed SC-gap anisotropy strongly indicates that the pairing interaction is not a conventional phonon-mediated isotropic one. 
Instead, the results suggest the importance of second-nearest-neighbor electronic interactions between the iron sites in the framework of $s_{\pm}$-wave superconductivity.

We would like to thank A. V. Chubukov, S. Maiti, R. Arita, and H. Kontani for valuable discussions and comments. This research is supported by JSPS through its FIRST Program.
%


\begin{thebibliography}{36}
\bibitem [{not({\natexlab{b}})}]{note_k}%
  \BibitemOpen
  \href@noop {} {} \bibinfo {note} {{The photoelectrons
  emitted from $E_F$ have a momentum parallel to the surface $k_{\parallel}$,
  given by $k_{\parallel}$ = $\sqrt{2m(E_{h\nu}-W)/\hbar^2}\sin\theta$ $\sim$
  0.512$\sqrt{(E_{h\nu}-W)\mathrm{[eV]}}\sin\theta$ $\mathrm{\AA}^{-1}$, where
  $m$, $\hbar$, $E_{h\nu}$, $W$, and $\theta$ are the electron mass, Planck's
  constant/2$\pi$, photon energy (= 6.994 eV), work function ($\sim$ 3.8 eV),
  and emission angle of photoelectrons, respectively. Since the photon energy
  is low in laser ARPES and the largest $\theta$ for our measurements was about
  20$^{\circ}$, the maximum $k_{\parallel}$ that can be probed is limited to
  about 0.3 $\mathrm{\AA}^{-1}$.}}\BibitemShut {Stop}%
\end{thebibliography}

%

\end{document}